\documentclass[10pt]{article}
\usepackage{multicol}
\usepackage{graphicx}
\usepackage{amsmath}

\begin{document}

\begin{center}
{\Large \bf Analysis of kinetic freeze out temperature and
transverse flow velocity in nucleus-nucleus and proton-proton
collisions at same center of mass energy}

\vskip1.0cm

M.~Waqas$^{1,}${\footnote{Corresponding author. Email (M. Waqas):
waqas\_phy313@yahoo.com}}, G. X. Peng$^{1,2,3}$
{\footnote{Corresponding author. Email (G. X. Peng):
gxpeng@ucas.ac.cn}}, Z. Wazir$^{4,}$ {\footnote{Z.Wazir:
zafar\_wazir@yahoo.com }},\\
Hai-Ling Lao$^{5,}$ {\footnote{H-L Lao: hailinglao@pku.edu.cn}}

\vskip.25cm

{\small\it $^1$ School of Nuclear Science and Technology,
University of Chinese Academy of Sciences, Beijing 100049, China

$^2$ Theoretical Physics Center for Science Facilities, Institute
of High Energy Physics, Beijing 100049, China

$^3$ Synergetic Innovation Center for Quantum Effects \&
Application, Hunan Normal University, Changsha 410081, China,

$^4$ Department of physics, Ghazi university, Dera ghazi khan,
Pakistan,

$^5$ Center for high energy physics, Peking university, Beijing
100871, China}
\end{center}

\vskip1.0cm

{\bf Abstract:} Transverse momentum spectra of {\bf different
types of} identified charged particles in central Gold-Gold
(Au-Au) collisions, and inelastic (INEL) or non-single-diffrative
(NSD) proton-proton (pp) collisions at the Relativistic Heavy Ion
Collider (RHIC), as well as in central and peripheral Lead-Lead
(Pb-Pb) collisions, and INEL or NSD pp collisions at the Large
Hadron Collider (LHC) are analyzed by the blast wave model with
Tsallis statistics. The model results are approximately in
agreement with the experimental data measured by STAR, PHENIX and
ALICE Collaborations in special transverse momentum ranges.
Kinetic freeze out temperature and transverse flow velocity are
extracted from the transverse momentum spectra of the particles.
It is shown that kinetic freeze out temperature of the emission
source depends on mass of the particles, which reveals the {\bf
mass differential kinetic freeze out scenario} in collisions at
RHIC and LHC. Furthermore, the kinetic freeze out temperature and
transverse flow velocity in central nucleus-nucleus (AA)
collisions are larger than in peripheral collisions, and both of
them are slightly larger  in peripheral nucleus-nucleus collisions
or almost equivalent to that in pp proton-proton collisions at the
same center of mass energy which shows their similar thermodynamic
nature.
\\



\vskip1.0cm
\begin{multicols}{2}

{\section{Introduction}}

It is a common sense that transition to quark matter happens in
special circumstances, e.g. astronomically at the core of a dense
compact star [1,2], and territorially in relativistic heavy-ion
collisions [3-5]. In later case, when the incident energy of
nucleus is in the order of GeV, a hot matter called quark-gluon
plasma (QGP) is produced. The QGP behaves like nearly a perfect
fluid and it expands very rapidly and emit large amount of
radiations which result in cooling of the matter. In fact, its
direct detection  and study is extremely difficult due to having a
very short life time, but the particles or radiations emission
from QGP are good objects to observe the QGP properties. One of
the most central concepts in thermodynamics and statistical
mechanics [6] is the temperature, which is very important in both
the thermal and subatomic physics, because it has extremely wide
applications in experimental measurements and theoretical studies.
Four types of different temperatures namely initial temperature
($T_\mathrm{i}$), chemical freeze out temperature
($T_\mathrm{ch}$), effective temperature ($T_\mathrm{eff}$) and
kinetic freeze out temperature ($T_0$) can be found in literature.
The detailed explanation of these temperatures is given in one of
our recent work [7]. $T_0$ describes the excitation degree of the
interacting system at the stage of thermal/kinetic freeze out. A
natural question arises that how many kinetic freeze out
temperatures ($T_0$) are there at the stage of thermal freeze out.
In general, both $T_0$ and $T_\mathrm{ch}$ happens simultaneously
or the former happens later. The behavior of $T_\mathrm{ch}$ is
studied in [8, 9] while $T_0$ has a more complex situation. There
are three main issues to be checked. It is known that $T_0$ {\bf
initially, increase sharply in central collisions from low energy
ranges upto almost 10 GeV in Beam Energy Scan (BES) energies} [10,
11], and then saturates upto 39 GeV. However after 39 GeV, this
tendency can be saturated, decrescent or incresecent. It is needed
to check the correct tendency. Secondly, $T_0$ can be slightly
larger in central collision, or slightly smaller than or
approximately equal to that in peripheral collision. It is needed
to check that which collisions has larger $T_0$. Thirdly, there
are different kinds of kinetic freeze out (KFO) scenarios [12--14]
available in literature, it is possible that $T_0$ can give
single, double, or multiple values for the emission of different
kinds of particles in various collisions. It is necessary to check
the correct freeze out scenario.

The solution  of all the above questions is very interesting but
difficult. Particularly, the study of excitation function of $T_0$
is required in the first issue which is already studied in our
recent work [11, 15]. The second issue is analyzed and studied in
[16]. It is also important to study $T_0$ in inelastic (INEL) or
non-diffractive (NSD) proton-proton (pp) collisions at RHIC and
LHC energies. In case of third issue, the $p_T$ spectra of
different types of particles can be used to extract $T_0$, along
with the accompanying results of $\beta_T$, which also exhibits a
complex situation as $T_0$. Numerous methods can be used for the
extraction of $T_0$ and $\beta_T$. In the present work, we shall
use the blast wave model with Tsallis statistics [12, 17] in order
to extract $T_0$ and $\beta_T$ from the $p_T$ spectra in central
and peripheral gold-gold (Au-Au)and lead-lead (Pb-Pb) collisions
as well as INEL or NSD pp collisions at RHIC and at LHC. The model
results are compared with the experimental data measured by STAR
and PHENIX collaborations. The remainder of the paper is discussed
in the sections below.
\\

{\section{The model and method}}

The two main processes of multi-particle production in high energy
collisions are soft and hard excitation process. Soft excitation
contributes in the production of most light flavor particles and
is distributed in a narrow $p_T$ range of less than 2-3 GeV/c or a
little more. Generally, due to being small $p_T$ fraction, the
hard scattering process does not contribute mainly to $T_0$ and
$\beta_T$. In case of the narrow $p_T$ range, we can neglect the
contribution of hard scattering process. However, it is expected
that the contribution fraction of hard scattering process increase
with increase of collision energy. Thus, the contribution of hard
scattering process cannot be neglected in collisions at very high
energy, though the soft excitation process is still the main
contributor of particles production.

The description of this process has many choices of formalism
which include but are not limited to the blast wave model with
Tsallis statistics [12], blast wave model with Boltzmann
statistics [18--20], Tsallis and related distributions with
various formalisms [21--25], the (multi-) Standard distribution
[26, 27], the Hagedorn thermal model [28], Erlang distribution
[29--31] and the schwinger mechanism [6,32--35]. As a simple and
practical application, in this work, we choose the blast-wave
model with Tsallis statistics [11, 12, 19] to be more convenient
due to the fact that the Tsallis and its alternative forms can be
used for the fit of wider spectra, and $T_0$ and $\beta_T$ can
easily be extracted by using the blast wave model.

According to refs. [11, 12, 19], the probability density function
of $p_T$ can be given by
\begin{align}
f_1(p_T)=&\frac{1}{N}\frac{\mathrm{d}N}{\mathrm{d}p_\mathrm{T}}=C  p_T m_T \int_{-\pi}^\pi d\phi\int_0^R rdr \nonumber\\
& \times\bigg\{{1+\frac{q-1}{T_0}} \bigg[m_T  \cosh(\rho)-p_T \sinh(\rho) \nonumber\\
& \times\cos(\phi)\bigg]\bigg\}^\frac{-1}{(q-1)}
\end{align}
where C is the normalization constant which results in the
integral of Eq. (1) to be normalized to 1,
$m_T=\sqrt{p_T^2+m_0^2}$ is the transverse mass, $m_0$ is the rest
mass of the particle, $\phi$ is the azimuthal angle, r is the
radial coordinate, R is the maximum r, q is the entropy index,
$\rho=\tanh^{-1} [\beta(r)]$ is the boost angle,
$\beta(r)=\beta_S(r/R)^{n_0}$ is a self-similar flow profile,
$\beta_S$ is the flow velocity on the surface, and $n_0=1$ is used
in the original form [12]. In particular, $\beta_T=(2/R^2)\int_0^R
r\beta(r)dr=2\beta_S/(n_0+2)=2\beta_S/3$. The hard scattering
process contributes the spectrum in wide or low+high $p_T$ range
that is described by the QCD calculus [36--38] or the Hagedorn
function [28], which is an inverse power law as shown in the
function below
\begin{align}
f_0(p_T)=\frac{1}{N}\frac{dN}{dp_T}= Ap_T \bigg( 1+\frac{p_T}{
p_0} \bigg)^{-n},
\end{align}
where A is the normalization constant that normalizes the integral
of Eq. (2) to unity, and $p_0$ and n are the free parameters.

If the contributions of both the soft excitation and hard
scattering processes are involved, the experimental $p_T$ spectrum
distributed in a wide range can be described by a superposition.
We have
\begin{align}
f_0(p_T)=\frac{1}{N}\frac{dN}{dp_T}=kf_S(p_T)+(1-k)f_H(p_T),
\end{align}
where $k$ and $(1-k)$ denotes the contribution fraction of the
soft excitation and hard scattering process. Naturally, the
integral of Eq. (3) is normalized to 1.

According to the Hagedorn model [28], the usual step function
$\theta(x)$ can also be used in order to superimpose the
contributions of soft excitation and hard scattering processes,
where $\theta(x)=0$ if $x<0$ and $\theta(x)=1$ if $x\geq0$. We
have
\begin{align}
f_0(p_T)&=\frac{1}{N}\frac{dN}{dp_T} \nonumber\\ &=A_1
\theta(p_1-p_T) f_S(p_T) + A_2 \theta(p_T-p_1)f_H(p_T),
\end{align}
where $A_1$ and $A_2$ are the normalization constants which
results in the two components to be equal to each other at $p_T$ =
$p_1$. The integral of Eq. (4) is normalized to 1. The
contribution fraction $k(1-k)$ of the soft excitation (hard
scattering) process is the integral of the first (second)
component.

Eq. (3) and (4) are two different superpositions. Eq. (3) exhibits
the contribution of soft component from 0 up to 2-3 GeV/c or a
little more, which is in the low-$p_T$ region. The hard component
contributes in the whole $p_T$ range, i.e. low+high $p_T$ region.
There is overlap for the two contributions in the low-$p_T$ region
due to Eq. (3). In Eq. (4), the soft component contributes from 0
to $p_1$, while from $p_1$ to maximum is the contribution of hard
component. There is no overlap for the two contributions due to
Eq.(4).

It should be noted that in the present work, we have used Eq. (1)
only, because we have used single component TBW model, but in case
if the single component of the model is not enough to describe the
spectra, then Eq. (3) and (4) can be used, and they are presented
in order to explain the whole methodology.

To extract $T_0$ and $\beta_T$ , the $p_T$ range studied in this
paper is not too wide. We thus do not need to consider the
contribution of hard component. Then, the second component in Eqs.
(3) and (4) can be neglected. Only the first component in the two
equations are necessary to consider, or in other words the two
equations degrade into one, i.e. Eq. (1). In the following
section, we shall only use Eq. (1) to fit the experimental data
and to extract $T_0$ and $\beta_T$. Before going to section 3, we
would like to make it clear that there are two ways of fit. (1)
Individual fit (2) Combined fit. The individual fit for each
particle specie is done in our recent work. If the parameters
obtained from the individual fit are similar, then we can say that
the combine fit is fine. The combined fit means that we have to
use the same parameters to fit various spectra in a very narrow
$p_T$ range e:g 1-2 GeV/c for $\pi$ and 1.5-2.6 GeV/c for $K$. The
combined fit can be used if different $p_T$ regions are used for
different spectra.
\\

{\section{Results and discussion}}

Figure 1 demonstrates the transverse momentum spectra ,
[(1/2$\pi$$p_T$) $d^2$N/dyd$p_T$] of identified charged particles
($\pi^-$,$K^-$ and $\bar p$) produced in (a)-(b) in the
mid-rapidity range $|y|-0.5<0$ at 62.4 GeV [39] and (c)-(d) at 200
GeV [40] with Pseudo-rapidity $|\eta|<0.35$ in Au-Au collisions.
Panel (e)-(f) corresponds to the spectra [(E$d^3$$\sigma$/$dp^3$]
for $\pi^-$,$K^-$ and $\bar p$ with $|\eta|<0.35$ produced in pp
collisions at 62.4 and 200 GeV [41] . The symbols are used to
represent the experimental data measured by STAR and PHENIX
[39--41] Collaborations at RHIC. The solid curves are the results
of our fits by using Eq.(1) and the corresponding result of their
data/fit are followed in each panel. Panels (a) and (c) correspond
to the central Au-Au collisions (0--10\% and 0--5\% centrality
respectively), while panel (b) and (d) correspond to the
peripheral Au-Au collisions (40--80\% and 60--92\% centrality
respectively).
The values of free parameters ($T_0$, $\beta_T$, q), normalization
constant ($N_0$), $\chi^2$, and degree of freedom (dof) are listed
in table 1.

\begin{figure*}[htbp]
\begin{center}
\includegraphics[width=16.cm]{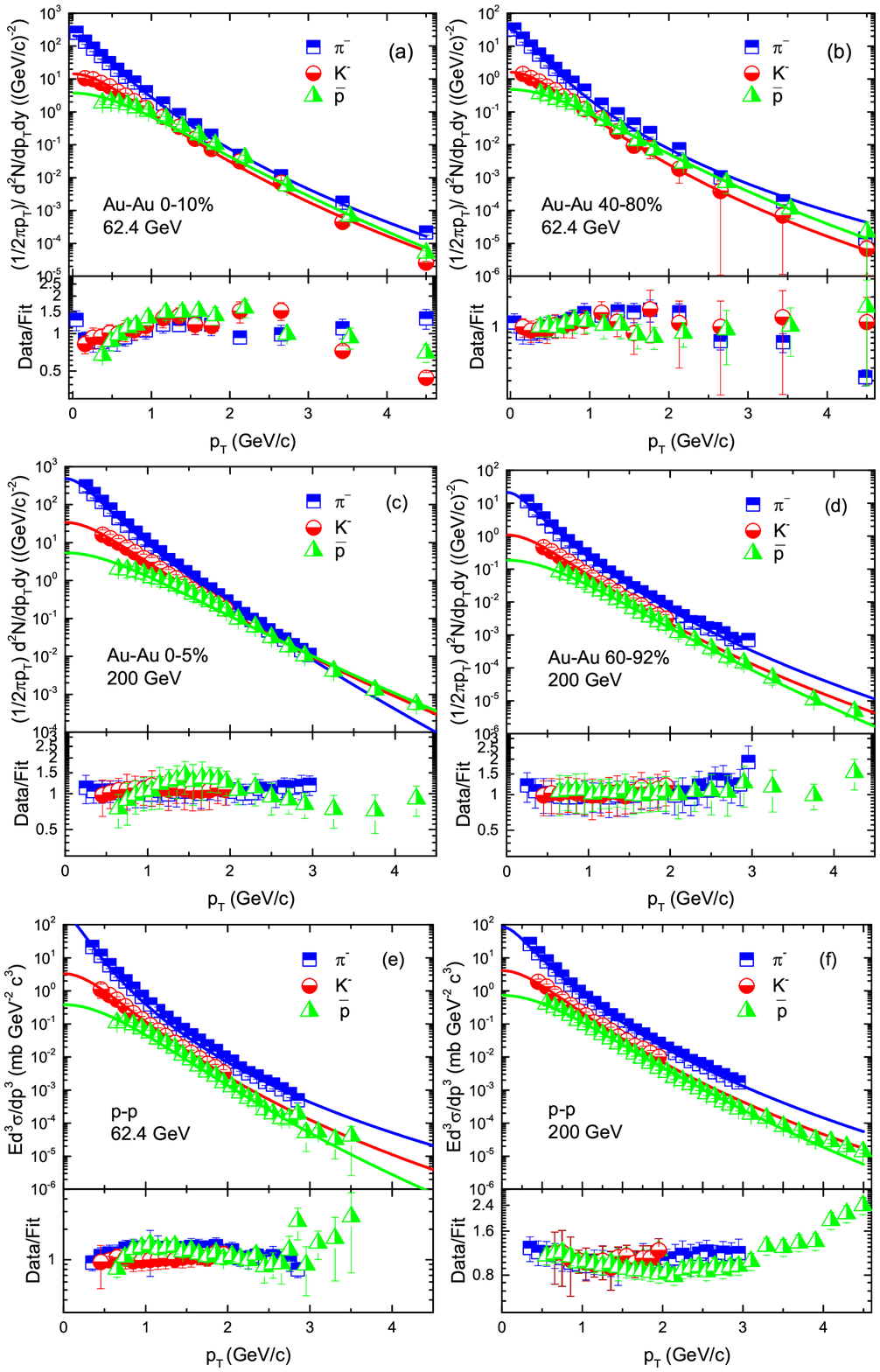}
\end{center}
Fig. 1. Transverse momentum spectra of $\pi^-$, $K^-$ and $\bar p$
produced in central and peripheral Au-Au collisions (a)-(b) at
62.4 GeV  in the mid-rapidity $|y|-0.5<0$ [39], (c)-(d) at 200 GeV
with $|\eta|<0.35$ [40] and pp collisions at 62.4 GeV in panel (e)
and 200 GeV in panel (f) in $|\eta|<0.35$ [41]. The symbols
represent the experimental data measured by STAR and PHENIX
collaborations [39, 40, 41] at RHIC, while the curves are our
fitted results by using the blast wave model with Tsallis
statistics, Eq (1). The corresponding results of data/fit are
presented in each panel.
\end{figure*}

\begin{table*}
{\scriptsize Table 1. Values of free parameters ($T_0$ and
$\beta_T$), entropy index (q), normalization constant ($N_0$),
$\chi^2$, and degree of freedom (dof) corresponding to the curves
in Figs. 1--2. \vspace{-.50cm}
\begin{center}
\begin{tabular}{ccccccccc}\\ \hline\hline
Figure  & Centrality & Particle  & $T_0$ (GeV & $\beta_T$ (c) & $q$   & $N_0$      &  $\chi^2$  & dof\\
\hline
Fig. 1  & 0--10\%   & $\pi^-$  & $0.072\pm0.005$ & $0.447\pm0.008$ &$1.074\pm0.05$   & $14\pm 2$      &  15  & 17\\
Au-Au   &          & $K^-$    & $0.082\pm0.007$ & $0.415\pm0.009$ & $1.062\pm0.06$  &$2.3\pm0.5$    &  142 & 16\\
62.4 GeV &         & $\bar p$ & $0.088\pm0.006$ & $0.411\pm0.009$ & $1.050\pm0.05$  & $1.2\pm0.3$   & 68   & 14\\
\cline{2-8}
        & 40--80\% & $\pi^-$  & $0.055\pm0.006$ & $0.423\pm0.011$ & $1.105\pm0.03$  & $1.5\pm0.35$   & 150 & 17\\
        &          & $K^-$    & $0.064\pm0.005$ & $0.390\pm0.012$ & $1.08\pm0.04$   &$2.3\pm0.6$    &  2  & 16\\
        &          & $\bar p$ & $0.071\pm0.006$ & $0.370\pm0.012$ & $1.07\pm0.03$   & $1.5\pm0.3$   & 3   & 14\\
\hline
Fig. 1  & 0--20\%  & $\pi^-$  & $0.094\pm0.006$ & $0.499\pm0.011$ & $1.025\pm0.05$    &$40\pm7$     & 3 & 28\\
Au-Au  &           & $K^-$    & $0.099\pm0.007$ & $0.452\pm0.012$  & $1.05\pm0.04$    &$6.62\pm1$     & 1 & 16\\
200 GeV &          & $\bar p$ & $0.104\pm0.005$ & $0.439\pm0.009$  & $1.046\pm0.05$   &$2.1\pm0.3$     & 12 & 22\\
\cline{2-8}
        & 60--88\% & $\pi^-$  & $0.070\pm0.005$ & $0.461\pm0.013$ & $1.07\pm0.05$  & $1.3\pm0.2$      &  5 & 28\\
        &          & $K^-$    & $0.077\pm0.005$ & $0.411\pm0.011$ & $1.067\pm0.04$ & $0.17\pm0.2$      & 1 & 16\\
        &          & $\bar p$ & $0.086\pm0.007$ & $0.390\pm0.008$ & $1.046\pm0.04$  & $0.055\pm0.007$     & 3 & 22\\
\hline
Fig. 1  & $-$      & $\pi^-$  & $0.040\pm0.005$ & $0.410\pm0.012$  & $1.1\pm0.07$     & $5.7\pm0.05$       & 21 & 26\\
$pp$    &          & $K^-$    & $0.056\pm0.006$ & $0.370\pm0.009$  & $1.078\pm0.06$   &$ 0.40\pm0.04$     & 1 & 16\\
62.4 GeV &          & $\bar p$ & $0.060\pm0.004$ & $0.352\pm0.011$ & $1.055\pm0.05$   &$0.09\pm0.05$      & 30 & 27\\
\hline
Fig. 1  & $-$      & $\pi^-$  & $0.065\pm0.007$ & $0.450\pm0.010$  & $1.08\pm0.07$     & $4.8\pm0.4$       & 5 & 27\\
$pp$    &          & $K^-$    & $0.071\pm0.005$ & $0.375\pm0.011$  & $1.078\pm0.05$    &$ 0.6\pm0.08$      & 4 & 16\\
200 GeV &          & $\bar p$ & $0.078\pm0.006$ & $0.345\pm0.013$  & $1.057\pm0.06$    &$0.2\pm0.03$       & 86 & 33\\
\hline
Fig. 2   & 0--5\% & $\pi^-$   & $0.108\pm0.007$ & $0.530\pm0.014$ & $1.016\pm0.04$ & $128.9\pm27$  & 89 & 40\\
Pb-Pb    &        & $K^-$     & $0.111\pm0.005$ & $0.505\pm0.012$ & $1.055\pm0.03$ & $ 17.12\pm3$  & 30 & 36\\
2.76 TeV &        & $\bar p$  & $0.116\pm0.005$ & $0.496\pm0.013$ & $1.044\pm0.02$ &  $4.9\pm1$     & 72 & 37\\
\cline{2-8}
 & 70--80\% & $\pi^-$         & $0.091\pm0.006$ & $0.497\pm0.012$ & $1.053\pm0.03$ & $1.1\pm0.3$      & 209 & 40\\
 &          & $K^-$           & $0.094\pm0.005$ & $0.480\pm0.010$ & $1.05\pm0.04$ & $0.14\pm0.04$  & 64 & 36\\
 &          & $\bar p$        & $0.101\pm0.006$ & $0.470\pm0.009$ & $1.035\pm0.04$ & $0.054\pm0.001$ & 62 & 37\\
\hline
Fig. 2   & 0--5\% & $\pi^-$             & $0.126\pm0.006$ & $0.550\pm0.010$ & $1.01\pm0.05$ &  $6.6\pm0.8$     & 756 & 35\\
Pb-Pb    &        & $K^-$               & $0.128\pm0.005$ & $0.539\pm0.014$ & $1.015\pm0.04$ & $1.2\pm0.04$  & 801  & 34\\
5.02 TeV &        & $\bar p$            & $0.131\pm0.007$ & $0.528\pm0.010$ & $1.002\pm0.04$ &  $3.5\pm0.4$   & 247 & 32\\
\cline{2-8}
 & 60--80\% & $\pi^-$             & $0.100\pm0.006$ & $0.522\pm0.013$ & $1.05\pm0.06$ & $4.2\pm0.4$  & 435 & 35\\
 &          & $K^-$                 & $0.113\pm0.005$ & $0.516\pm0.009$ & $1.04\pm0.05$ & $0.90\pm0.04$   & 73 & 34\\
 &          & $\bar p$                & $0.121\pm0.007$ & $0.502\pm0.012$ & $1.015\pm0.04$ &  $6.5\pm0.3$  & 261 & 32\\
\hline
Fig. 2 & $-$ & ($\pi^++\pi^-$)/2 & $0.087\pm0.006$ & $0.486\pm0.008$ & $1.046\pm0.03$ & $0.35\pm0.06$   & 385  & 22\\
$pp$   &     & ($K^++K^-$)/2     & $0.091\pm0.007$ & $0.468\pm0.012$ & $1.04\pm0.04$ &  $0.048\pm0.005$ & 8 & 17\\
2.76 TeV  &     & ($p+\bar p$)/2    & $0.097\pm0.005$ & $0.448\pm0.011$ & $1.03\pm0.04$ & $0.02\pm0.007$ & 13 & 27\\
\hline
Fig. 2 & $-$ & $\pi^-$ & $0.094\pm0.006$ & $0.513\pm0.013$ & $1.13\pm0.05$ & $1.6\pm0.3$   & 235  & 35\\
$pp$   &     & $K^-$     & $0.098\pm0.007$ & $0.491\pm0.012$ & $1.12\pm0.07$ &  $0.34\pm0.06$ & 904 & 33\\
5.02 TeV  &     & $\bar p$    & $0.103\pm0.005$ & $0.481\pm0.011$ & $1.096\pm0.05$ & $0.16\pm0.04$ & 412 & 31\\
\hline
\end{tabular}%
\end{center}}
\end{table*}

Figure 2 is similar to figure 1, but it shows the transverse
momentum spectra of $\pi^-$, $K^-$ and $\bar p$ in panel (a)-(b)
and in (c)-(d) with $|y|<0.5$ in Pb-Pb collisions at 2.76 [42] and
5.02 TeV [43] respectively. Panels (e) and (f) represents the
transverse momentum spectra of ($\pi^++\pi^-$)/2, ($K^++K^-$)/2
and ($p+\bar p$)/2 with $|y|<1$ at 2.76 TeV [44] and $\pi^-$,
$K^-$ and $\bar p$ with $|y|<0.5$ at 5.02 TeV [43] respectively.
The spectra of $\pi^-$, $K^-$and $\bar p$ for Pb-Pb central
collisions at 2.76 TeV are scaled by the factor $2^9$. The curves
are the result of our fitting by using Eq. (1), while each panel
is followed by the result of its data/fit. The experimental data
measured by the ALICE Collaboration [42, 43, 44] at LHC are
represented by the symbols.  Panels (a) and (c) corresponds to the
central Pb-Pb collisions (0--5\% centrality), while panel (b) and
(d) corresponds to the peripheral Pb-Pb collisions (80--90\%
centrality). $N_{ev}$ on the vertical axis represents the number
of events. In figure 1 and 2, it can be seen that the model
results describe approximately the experimental data in special
$p_T$ ranges at RHIC and LHC. It is noteworthy that in the fitting
of some spectra, the $p_T$ range is above 0.4 GeV/c due to the
unavailability of $p_T$ range less than 0.4 GeV/c.

\begin{figure*}[htbp]
\begin{center}
\includegraphics[width=16.cm]{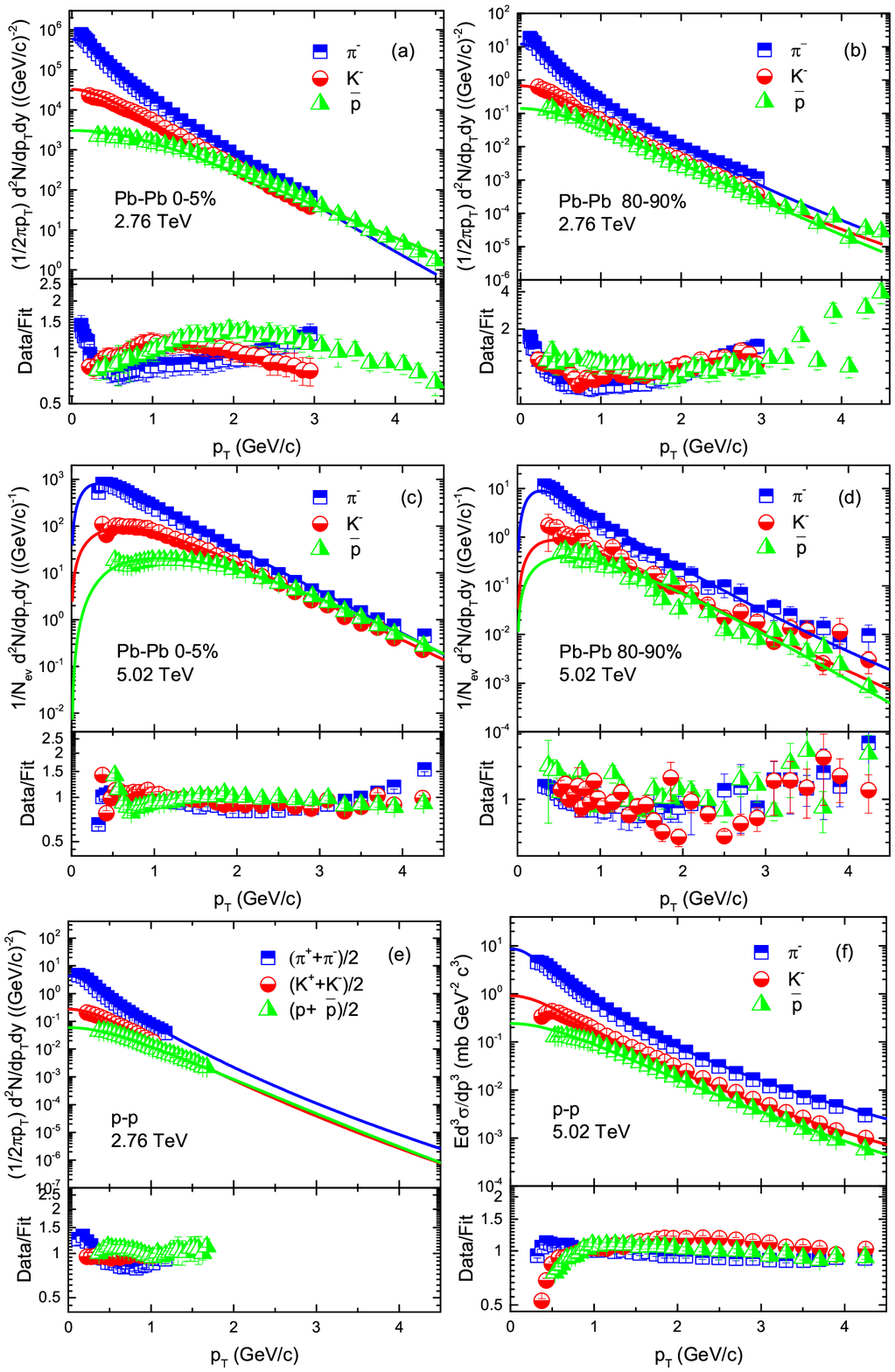}
\end{center}
Fig. 2. is similar to fig. 1, but shows the spectra of the
particles in Pb-Pb collision from panel (a)-(d), as well as in pp
collisions in panels (e)-(f). Panel (a)-(d): For $\pi^-$, $K^-$
and $\bar p$ with $|y|<0.5$ at 2.76 TeV [42] and 5.02 TeV [43]
respectively measured by ALICE Collaboration. Panel (e) and (f):
For ($\pi^++\pi^-$)/2, ($K^++K^-$)/2 and ($p+\bar p$)/2 with
$|y|<0.5$ at 2.76 TeV [44] and $\pi^-$, $K^-$ and $\bar p$ with
$|y|<0.5$ at 5.02 TeV [43] measured by ALICE Collaboration
respectively. The curves are the results of our fitting by using
the blast wave model with Tsallis statistics, Eq (1). Each panel
is followed by the results of their data/fit. The spectra in the
panels (a)-(b) for the central collisions are scaled by the factor
of $2^9$.
\end{figure*}

\begin{figure*}[htbp]
\begin{center}
\includegraphics[width=16.cm]{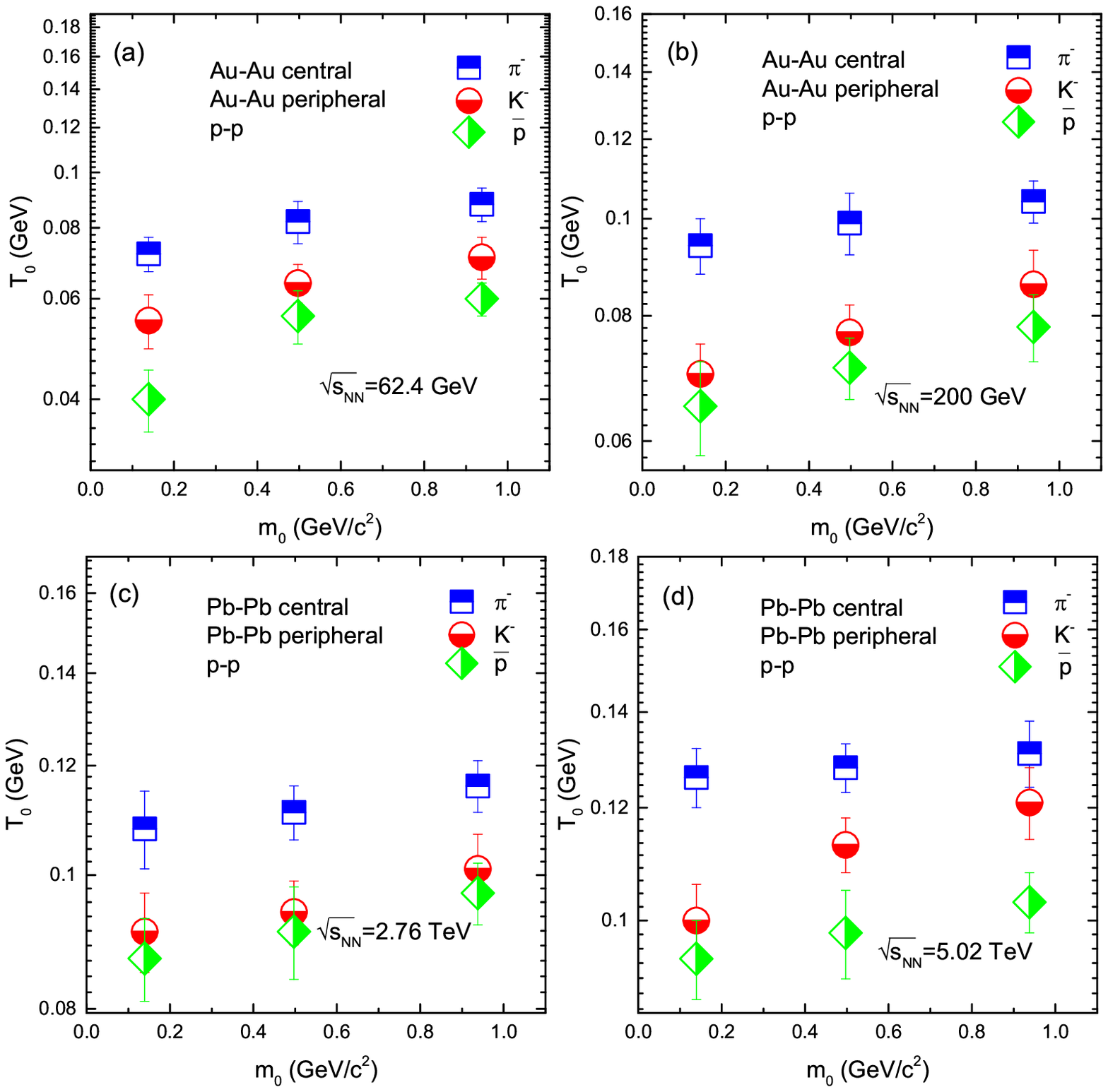}
\end{center}
Fig. 3. (a)-(b): Dependence of $T_0$ on $m_0$ in Au-Au central and
peripheral as well as in pp collisions at 62.4 GeV and 200 GeV
respectively. (c)-(d)dependence $T_0$ on $m_0$ in Pb-Pb central
and peripheral as well as in pp collisions at 2.76 TeV and 5.02
TeV respectively.
\end{figure*}

Figure 3 shows the change in trends of the kinetic freeze out
temperature ($T_0$). Panel (a)-(b) show the dependence of $T_0$ on
the rest mass of the particle in central and peripheral Au-Au as
well as in pp collisions at 62.4 GeV and 200 GeV respectively,
While panel (c)-(d) show the dependence of $T_0$ on the rest mass
of the particle in central and peripheral Pb-Pb as well in pp
collisions at 2.76 TeV and 5.02 TeV respectively. Different
symbols represent the values of parameter ($T_0$) for different
particles in different collisions. One can see that the values of
$T_0$ for heavier particles are larger than that for the lighter
particles, which show the early freeze out of the heavier
particles compared to the lighter particles. The different freeze
out for different particles reveal the mass differential freeze
out (multiple freeze out) scenario. Furthermore, $T_0$ in the
central nucleus-nucleus (AA) collisions is larger than in
peripheral collisions due to the reason that central collision
involves more participant nucleons in interaction, so the
collision is more violent and the system gets higher degree of
excitation, while the peripheral collision involves less
participants and attains low degree of excitation in the
interaction which results in smaller $T_0$. Larger $T_0$ in
central collisions indicates quicker approach of equilibrium of
the system. $T_0$ is also analyzed in pp collisions at RHIC and
LHC energies. It is observed that $T_0$ in peripheral AA
collisions is slightly larger than in pp collisions at the same
center of mass energy due to the involvement of more participant
nucleons in former interaction as compared to the later, or it is
approximately equal in both these collisions at the same center of
mass energy due to a not violent interaction and the later
statement indicates almost the same thermodynamic nature of the
parameters in the corresponding collisions of the same center of
mass energy (per nucleon pair). It is also observed that $T_0$ in
Pb-Pb collisions is larger than in Au-Au collisions, and in AA
collisions, it is larger than in pp collisions, which may indicate
its dependence on the size of interacting system. The present work
also shows the energy dependence of in AA and pp collisions. $T_0$
dependence on energy and on the size of the interacting system (in
AA or pp collisions) is a huge project which requires a lot of
data, therefore we will keep this project in future consideration.

\begin{figure*}[htbp]
\begin{center}
\includegraphics[width=16.cm]{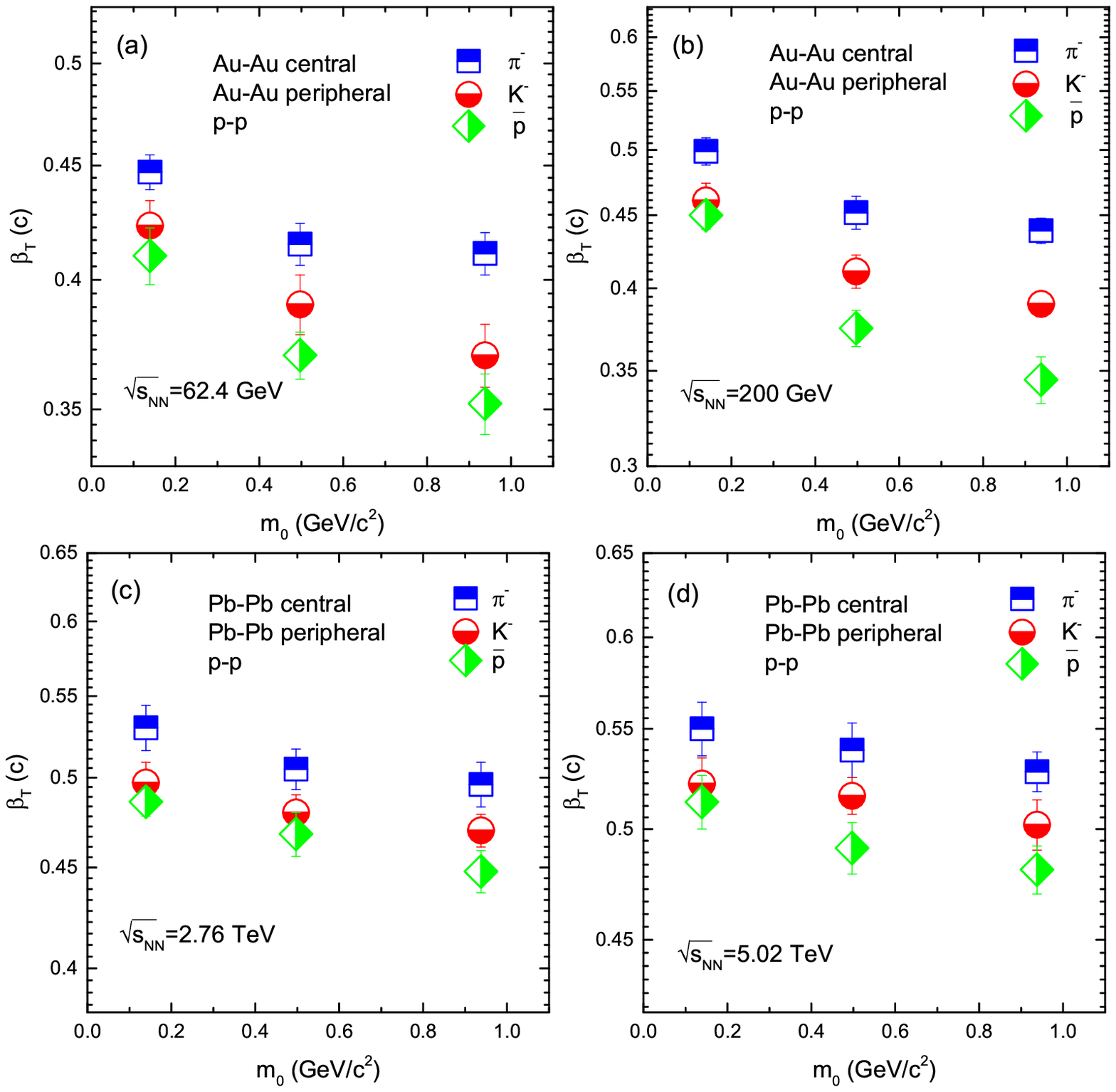}
\end{center}
Fig. 4. (a)-(b): Dependence of $\beta_T$ on $m_0$ in Au-Au central
and peripheral as well as in pp collisions at 62.4 GeV and 200 GeV
respectively. (c)-(d) dependence $\beta_T$ on $m_0$ in Pb-Pb
central and peripheral as well as in pp collisions at 2.76 TeV and
5.02 TeV respectively.
\end{figure*}

The transverse flow velocity ($\beta_T$) dependence on $m_0$ and
centrality is shown in figure 4. It can be obviously seen that
$\beta_T$ in Au-Au and Pb-Pb central collisions is larger than in
peripheral collisions, due to the involvement of large number of
participants in the interaction which results in more energy
deposition in the central collisions and the fireball expands more
rapidly. The heavier particles are observed to have smaller values
of $\beta_T$ as compared to the light particles. Furthermore,
$\beta_T$ in AA peripheral collisions is slightly larger than or
almost equivalent to pp collisions at the same center of mass
energy (per nucleon pair), and the collisions in both the
peripheral AA and pp interaction are not violent due to the
involvement of few nucleons in interaction and the fireball does
not expand quickly. Like $T_0$ in figure 3, $\beta_T$ in figure 4
is increasing with energy and also with the size of the
interacting system.

\begin{figure*}[htbp]
\begin{center}
\includegraphics[width=16.cm]{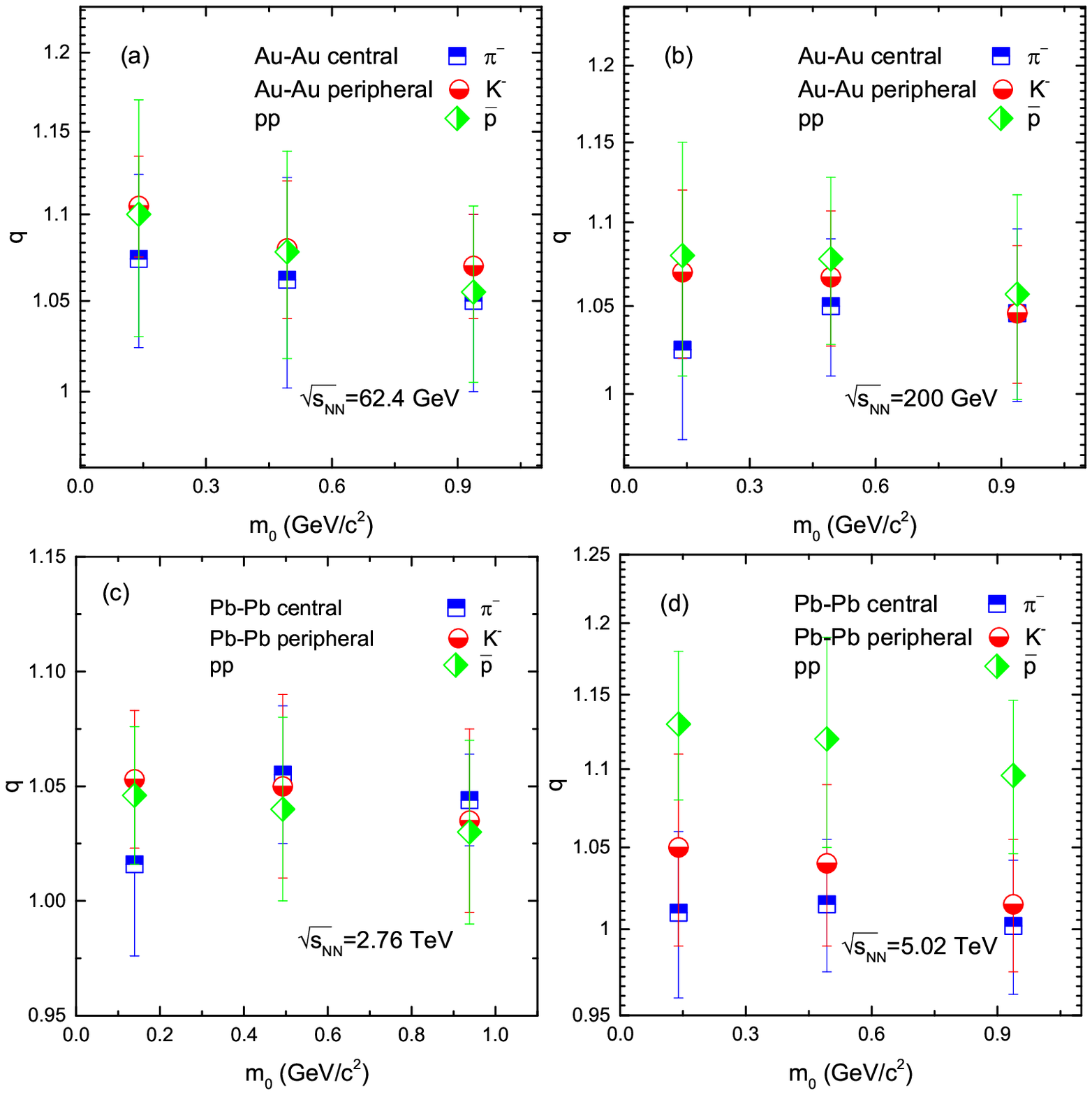}
\end{center}
Fig. 5. (a)-(b): Dependence of $q$ on $m_0$ and centrality in
Au-Au central and peripheral as well as in pp collisions at 62.4
GeV and 200 GeV respectively. (c)-(d) dependence $q$ on $m_0$ and
centrality in Pb-Pb central and peripheral as well as in pp
collisions at 2.76 TeV and 5.02 TeV respectively.
\end{figure*}

Figure 5 represents the dependence of the entropy index ($q$) on
$m_0$ and centrality. Different symbols and different colors
represent different collisions and different particles. From fig.
5(a)-(d), one can see that  in most cases, q is larger for the
lighter particles compared to heavier particles. e;g pion is
larger than the rest of two in most cases which shows that the
production of pion is more polygenetic than the others.
Additionally, $q$ is larger in the peripheral and pp collisions
compared to the central collisions which indicates that the
central nucleus-nucleus collisions has a quick approach to
equilibrium state as compared to the peripheral nucleus-nucleus
and pp collisions.

We would also like to point out that the values of normalization
constant ($N_0$) in table 1 are mass dependent, and $N_0$ decrease
with increasing the rest mass of the particle.

 Before going to the conclusions, we would like to give more
explanation of $p_T$ spectra and the parameters extracted from it.
The structure of $p_T$ spectra of the particles in high energy
collisions is very complex. Except the soft and hard $p_T$
regions, there also exists the very soft and very hard $p_T$
regions. The detail of the four $p_T$ regions can be found briefly
in [45]. The $p_T$ region of $p_T$ $<$ 0.2-0.3 is considered to be
the very soft region. The very soft $p_T$ region contributes to
the production of resonance which changes the slopes of the $p_T$
spectra and affects the values of free parameters. For-example,
the large contribution of resonance tends to reduce $T_0$ and
increase $\beta_T$ and this idea is giving strength to our current
conclusion of the multiple kinetic freeze out scenario. Among the
considered particles, pion is largely influenced by the resonance
[46, 47] and proton is more stable [42]. However, at higher $p_T$,
the hagedorn function contributes and it will not effect $T_0$ and
$\beta_T$.

The situation of $T_0$ and/or $\beta_T$ are very complex on the
basis of their dependence on energy and/or centrality. Different
models are analyzed in [48] which shows different results of $T_0$
and $\beta_T$. In addition, in [49], from RHIC to LHC, $T_0$
increase $\sim$9$\%$ and $\beta_T$ increase $\sim$65$\%$. However,
from RHIC to LHC, $T_0$ decrease $\sim$5$\%$ while an increase
$\sim$20$\%$ in $\beta_T$  from 39 to 200 GeV can be seen in [50,
51]. In [52, 53] there is no obvious change observed in $T_0$ from
RHIC to LHC, however $\beta_T$ shows an increase of $\sim$10$\%$.
Our recent results are similar to the above mentioned works
although the concrete values are not the same, especially, at
least one can observe the increasing excitation function of
$\beta_T$ with the increasing energy. In addition, $T_0$ and
$\beta_T$ show different trend from  central to peripheral
collisions in different literatures. In [10, 54, 55], $T_0$ is
increasing from central to peripheral collisions. Our present work
is inconsistent with [10, 54, 55], but consistent with [15, 45,
56] which gives larger $T_0$ in central collisions. Both the above
mentioned school of thoughts about the trend $T_0$ has their own
explanations. The present work explains the higher degree of
excitation of the interacting system in the central collisions due
to large number of participants involved in the interaction and
more energy is stored in the system. On the other hand, the larger
$T_0$ in the peripheral collisions explain the longer lifetime of
the fireball in the central collisions. It is noteworthy that the
present work results for $T_0$ are not consistent with [10, 54,
55] due to different methods and model with different conditions
and limitations. Even by using the same model with different
method and different conditions and limitations, we can get the
different results, e:g in our recent work [15] and in [10, 54, 55]
the blast wave model with boltzmann Gibbs statistics with
different conditions and limitations is used and the result is
different.

 It should be noted that Blast wave model with boltzmann Gibbs statistics
 (BGBW) has been extensively used for the description of the produced system
 at thermal/kinetic freezeout temperature. It is assumed by the BGBW model
 that the produced system has reached to thermal equilibrium so that a Boltzmann
 distribution with a radial profile can be used for the description of $p_T$
 spectra [18]. Nonetheless, the very limited low $p_T$ spectra can be described
 by the equilibrium distribution and is sensitive to the choice of specific $p_T$
 spectra (cover narrow $p_T$ range). Later, the Tsallis statistics was introduced
 to describe the particle production for an extended $p_T$ range (up to 5 GeV/c)
 in high energy collisions [57, 58, 59, 60, 61, 62]. The advantage of Tsallis
 statistics is that it introduces a new parameter "q" that describes the degree
 of non-equilibrium in the system, which is helpful in pp collisions [19] and AA
 peripheral collisions specially. The parameter q=1 in BGBW, while in TBW model
 q$<$1.25, which affects the $p_T$ range but it is not responsible to change the
 trend of the results.

It should be noted that there exists an entanglement in the
extraction of $T_0$ and $\beta_T$. In deed, if for central
collisions, one use a smaller $T_0$ and a larger $\beta_T$, a
decreasing trend for $T_0$ from peripheral to central collisions
can be obtained. At the same time, a negative correlation between
$T_0$ and $\beta_T$ will also be obtained. Thus, this situation is
in agreement with some references [10, 42, 55, 64, 65], but if for
central collisions, one even uses an almost unchanging or slightly
larger $T_0$ and a properly larger $\beta_T$, an almost invariant
or slightly increasing trend for $T_0$ from peripheral to central
collisions can be obtained [7, 15, 56, 63]. In order to show the
flexibility in the extraction of $T_0$ and $\beta_T$ , this work
has reported a decreasing trend for $T_0$ from central to
peripheral collisions, and a positive correlation between $T_0$
and $\beta_T$.

 In addition, we would point out that the present work is in
agreement as well as in disagreement with some works. This is due
to different models or methods as discussed before. In the present
work we have used the least square method with different
conditions. We have used slightly lager $T_0$ and $\beta_T$ in
central collisions and obtained larger $T_0$ in central collisions
as compared to peripheral collisions, and the same conditions with
the same method is used in [15, 45, 56]. Furthermore, the present
work used $n_0$=1 which is in close resemblance with hydrodynamic
profile as mentioned in [66]. $n_0$=2 is also a close
approximation to the hydrodynamic profile according to reference
[18]. $n_0$=1 or 2 does not effect the curve as well as the free
parameters. If one consider the quick decay of $\beta_r$ from the
surface to the center of emission source, one should use $n_0$=2,
but if the $\beta_r$ decay from the surface to the center emission
source is not quick, then $n_0$=1 will be used. Anyways, we did
not regard $n_0$ as a free parameter which is too inconsistent and
arguable in our opinion and it can have an effect on curve as well
as on the free parameters. According to [67], apart from the fact
that $n_0$ is mutable (from $0.0\pm10.1$ to $4.3\pm1.7$), but the
$p_T$ coverage is also narrow and particle dependent
($p_T$$\approx$0.20-0.70 GeV/c, 0.25-0.75 GeV/c and 0.35-1.15
GeV/c for $\pi^+$, $K^+$ and p respectively), that uses a single
kinetic freeze out scenario and give different result of $T_0$ and
$\beta_T$ from this work. If $n_0$ is regarded as a free parameter
and the particle dependent and narrow $p_T$ coverage is used,
similar result with [67] can be obtained.

In fact, many uncertainties arise from the fit function choice as
well as from the flow profile and also from the well-known
ambiguity in the fit results. It is always possible to trade $T_0$
against $\beta_T$ in a single $p_T$ spectrum. That is for a given
$p_T$ spectrum, there is a negative correlation among $T_0$ and
$\beta_T$. It is possible that we may obtain a positive or
negative correlation if we use a suitable $T_0$ and $\beta_T$ for
a set of $p_T$ spectra. In fact, there is an influence if a
changeable $p_T$ and /or $n_0$ choice on the extraction of the two
parameters are used. In our opinion, if we use a fixed flow
profile ($n_0$) and wide and fixed $p_T$ coverage for various
particles, the uncertainties can be reduced.
\\

{\section{Conclusions}}

The main observations and conclusions are summarized here.

(a) The transverse momentum spectra of the identified charged
particles [($\pi^-$, $K^-$ and $\bar p$) or($\pi^++\pi^-$,
$K^++K^-$ and $p+\bar p$)] produced in central and peripheral
Au-Au and Pb-Pb collisions as well as in pp collisions at RHIC and
LHC energy have been analyzed by TBW model. The model results are
in agreement with the experimental data in the special $p_T$ range
measured by STAR, PHENIX, and ALICE collaborations.

(b) Kinetic freeze out temperature and transverse flow velocity
are extracted from the transverse momentum spectra fitting. Both
$T_0$ and $\beta_T$ are observed to be larger in central
collisions than the peripheral collisions, and the results for
$T_0$ and $\beta_T$ at LHC are observed to be larger than that at
RHIC, which shows their dependence on the collision energy.

(c) $T_0$ and $\beta_T$ are mass dependent, $T_0$ increase with
increase of $m_0$ which reveals the multiple kinetic freeze out
scenario, however $\beta_T$ decrease with $m_0$.

(d) The kinetic freeze out temperature and transverse flow
velocity in peripheral Au-Au collisions and pp collisions at 62.4
and 200 GeV as well as in peripheral Pb-Pb collisions and pp
collisions at 2.76 and 5.02 TeV are similar and have similar trend
which exhibits the similar thermodynamic nature of the parameters
in peripheral AA and pp collisions at the same center of mass
energy (per nucleon pair).
\\

{\bf Acknowledgments}

The authors would like to thank support from the National Natural
Science Foundation of China (Grant Nos. 11875052, 11575190, and
11135011).

\end{multicols}
\end{document}